\documentstyle[twoside,fleqn,espcrc2,epsfig,psfig]{article}

\newcommand{\AmS}{{\protect\the\textfont2

  A\kern-.1667em\lower.5ex\hbox{M}\kern-.125emS}}

\title{Percolation and Deconfinement in SU(2) Gauge Theory
\thanks{The work has been supported by the TMR network ERBFMRX-CT-970122
and the DFG under grant Ka 1198/4-1.} 
}

\author{S. Fortunato$^{\mathrm a}$ and H. Satz
\address{Fakult{\"a}t f{\"u}r Physik, 
    Universit{\"a}t Bielefeld,\\
    D-33501 Bielefeld, Germany}} 
       
\begin{document}

\begin{abstract}
We show that deconfinement in SU(2) gauge theory can be
described by the percolation of site-bond clusters of like-sign Polyakov
loops. In particular, we find that in 2+1 dimensions
the percolation variables coincide with those of 
the 2-dimensional Ising model.

\end{abstract}

\maketitle

\section{INTRODUCTION}

    The critical behaviour of the Ising model can today be
    reformulated in terms of percolation theory: magnetization sets in
    when suitably defined clusters of parallel spins reach the dimensions
    of the system \cite{Co80}. 
    In particular, the critical exponents for percolation
    then become equal to the Ising exponents.
    
    We extend this description of critical behaviour in terms of
    percolation to the deconfinement transition in SU(2) gauge theory 
    for two space dimensions. We show that the percolation of Polyakov loop
    clusters (taken to be suitably defined 
    areas of Polyakov loops $L$ of the same
    sign) leads to the correct deconfinement temperature and to the correct
    critical exponents for the deconfinement.
  
    In contrast to the conventional studies of $L(T)$, the use of percolation
    strenght $P(T)$ as deconfinement order parameter remains possible also in
    the presence of dynamical quarks and could thus constitute a genuine
    deconfinement order parameter for full QCD.

\section{PERCOLATION AND CRITICAL \\
    PHENOMENA}

    The percolation problem \cite{Stauf94} is easy to formulate: 
    just place randomly pawns on a chessboard.
    Regions of adjacent pawns form 
    {\it clusters}.  Percolation theory deals with
    the properties of these clusters when the chessboard is infinitely
    large. If one of the clusters spans the chessboard from 
    one side to the opposite one, we say that the cluster 
\noindent{\it percolates}. 


    Quantitatively, one counts how many pawns belong 
    to each cluster and calculates two quantities:\\
      {$\bullet$}  The {\it average cluster
        size } S, defined as:
      {
        \begin{eqnarray}
          S={\sum_{s} \Bigg({{n_{s}s^2}\over{\sum_{s}{n_{s}s}}}\Bigg)}~.
        \end{eqnarray}}
      
      Here $n_{s}$ is the number of clusters of size $s$
      and the sums exclude the percolating cluster; this number 
      indicates how big on average the clusters are which do not
      percolate.\\
      {$\bullet$} The {\it percolation strength } P, 
      defined as:
      {
        \begin{eqnarray}{
            P=\frac{\mbox{\it size of the percolating cluster}}
            {\mbox{\it no. of lattice sites}}}~.
        \end{eqnarray} }
      \noindent 

      By varying the density of our pawns, a kind of phase transition occurs. 
      We pass from a phase of
      non-percolation to a phase in which one of the clusters percolates.
      The percolation strength P is the {\it order parameter}  of this
      transition: it is zero in the non-percolation phase and is different from
      zero in the percolation phase. 
      It is particularly interesting to study 
      what happens near the concentration
      $p_c$ where for the first time a percolating
      cluster is formed. 
      
      It turns out that P and S as functions of the density
      behave respectively 
      like the magnetisation M of the Ising model 
      and its magnetic susceptibility $\chi$  
      as functions of the temperature T (Table 1).

\newpage

\begin{table}[hbt]
\newlength{\digitwidth} \settowidth{\digitwidth}{\rm 0}
\catcode`?=\active \def?{\kern\digitwidth}
\caption{Behaviour around the critical point}
\label{tab:comp}
\begin{tabular*}{75mm}{@{}l@{\extracolsep{\fill}}ll}
\hline
 {Percolation}                  & {Ising Model}\\
\hline
\raisebox{-1.mm}{$P{\propto}(p-p_c)^{\beta}$ \makebox[30mm][r]{}}&
         \raisebox{-1.mm} {$M{\propto}(T_c-T)^{\beta}$}  \\ 
            \raisebox{-0.5mm}{\makebox[25mm][r]{{\it for} $p>p_c$}} 
          &   \raisebox{-0.5mm}{\makebox[25mm][r]{{\it for} $T<T_c$}}   \\
        \raisebox{-2.mm}{$S {\propto}|p-p_c|^{-\gamma}$}  &
        \raisebox{-2.mm}{${\chi} {\propto} |T_c-T|^{-\gamma}$}
\raisebox{-3.mm}{\vphantom{p} }   \\
\hline
        \end{tabular*}

\end{table}

\vspace{-8mm}
\section{DROPLET MODEL}

    The analogy of percolation with second order thermal phase transitions
    leads to a natural question:

    \vskip 2mm
    \centerline{Is it possible to describe these transitions} 
    \centerline{in terms
      of  percolation ?}

    \vskip 2mm
    The answer to that is given by the {\it droplet model},
    which reproduces precisely the results of the Ising model. 

    The droplet model establishes the following correspondence:
\[
\begin{array}{rcl}
\mbox{spins up (down)} & \Longrightarrow &
\mbox{occupied sites}\\
              \mbox{spins down (up)}& \Longrightarrow &
          \mbox{empty sites}\\
           \hspace{1.5mm}\begin{array}{r}
          \mbox{spontaneous}\\ \raisebox{.6mm}{\mbox{magnetization M}}
\end{array}
& \Longrightarrow& 
     \hspace{-1.2mm}\begin{array}{l}\mbox{strength of the}\\ 
\raisebox{.6mm}{\mbox{perc. cluster P}}
\end{array}
 \\
         \mbox{susceptibility $\chi$}& \Longrightarrow & 
         \mbox{av. cluster size S}\\
 \end{array}
\]

    The clusters are basically magnetic domains (either with positive or with
    negative magnetization). 
    When there is an infinite cluster (percolation),
    magnetization becomes a global property of the system.

    One has then to find at which temperature 
    percolation occurs and determine the corresponding critical exponents. 
    But this requires one further conceptual
    feature. If the 
    droplets are defined as clusters of  
    nearest-neighbour spins of the same type  
    (pure site percolation), then they do not lead to the correct Ising 
    results, 
    neither in two 
    dimensions, where the critical points coincide but not the 
    exponents \cite{Syk76}, nor in three dimensions, where even the thresholds
    are different \cite{Krum74}. In other words, correlations in 
    pure site percolation differ from thermal Ising correlations.

    To solve the problem, Coniglio and Klein \cite{Co80} introduced 
    a bond probability  $CK=1-e^{-2J{\beta}}$ 
    (J is the spin-spin coupling of the Ising model, $\beta$=1/kT, being k 
    the Boltzmann
    constant). The term $2J$ is just the difference $\Delta E$
    of energy that we have when we pass from 
    a pair of next-neighbour spins to a one 
    in which one of them is flipped. 
    To define the new "droplets" one has to 
    connect the parallel nearest-neighbour spins in 
    a cluster with a bond probability CK.
    Since CK is less than one, not all the parallel spins in a cluster become 
    part of the same droplet.

    With the Coniglio-Klein definition, the droplet model 
    reproduces the results of the Ising model, both in two and 
    in three dimensions.


\section{EXTENDING PERCOLATION TO SU(2) GAUGE THEORY}

    As it is well known, considerations of symmetry led 
    to the conjecture that SU(2) Gauge Theory
    and the Ising model belong to the same {\it universality class},
    that is they have the same critical exponents \cite{Svet82}.
    This close relation inspired our work. 

    If we take a typical SU(2) configuration at a certain temperature, there
    will be areas where L takes negative values, and areas where L takes 
    positive values. Both the positive and the negative 'islands' can be 
    seen as local regions of deconfinement. But as long as there are finite 
    islands of both signs, deconfinement remains a local phenomenon and the
    whole
    system is in the confined phase. When one of this islands percolates, that
    is it
    becomes infinite, then we can talk of deconfinement as a global phase of the 
    system.

    The main point of our work is to look for a suitable definition of droplets
    for SU(2). We expect our definition to be similar to the 
    Coniglio-Klein one. The problem is then to find the right 
    bond probability.
    In contrast to the Ising model, in which the spins can only take
    two values, we have now for L a continuous range of values. 
    Therefore, if we try to 
    build a bond probability like $CK=1-e^{-{\beta}{\Delta E}}$, 
    $\Delta E$ is not the same at any lattice site. 
    Because of that we used a local factor
    {
      \begin{equation}
        1-e^{-2{{\beta}_{eff}}L_iL_j}.
      \end{equation}}

    Here ${\beta}_{eff}$ is an effective coupling which was shown 
    in certain cases to 
    approximate SU(2) pure gauge theory as a system of nearest-neighbour interacting 
    spins \cite{Kar84}.\\

\vspace{-3mm}
\section{RESULTS}

    We performed simulations of SU(2) gauge theory 
    in 2+1 dimensions 
    (2 space dimensions and 1 time dimension)
    using four different 
    lattice sizes, $64^2$, $96^2$, $128^2$, $160^2$. 
    We focused on the case of $N_{\tau}=2$ lattice spacings
    in the time direction.
\vspace{-7mm}

     \begin{figure}[htb]
       \epsfig{file=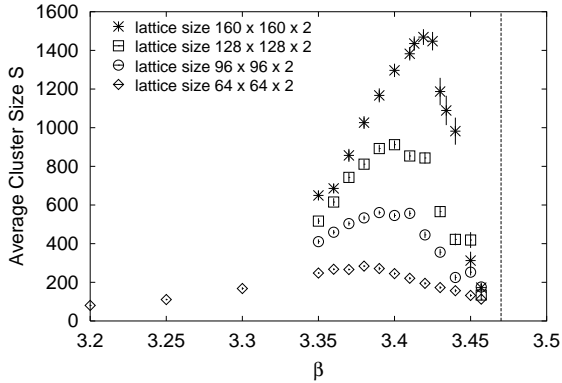,width=75mm}
\vspace{-12mm}
      \caption{ Average Cluster Size S of our SU(2)
        droplets for the four lattice sizes we used. The peaks approach
        the physical threshold (dotted line) for larger lattices.}
       \label{Scomp}
     \end{figure}

\vspace{-7mm}

\begin{figure}[htb]
       \vspace{5pt}
       \epsfig{file=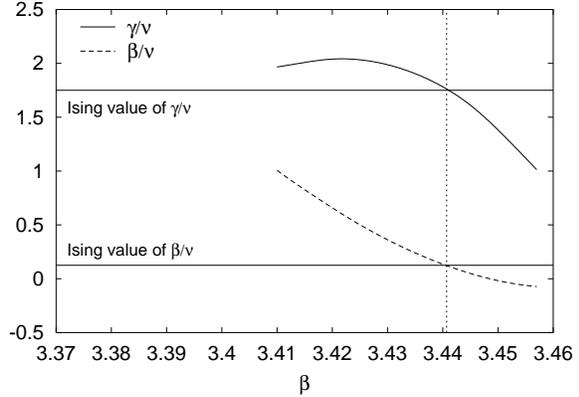,width=75mm}
\vspace{-12mm}
       \caption{ Best fit values for the exponents' ratios ${\beta}/{\nu}$
         and ${\gamma}/{\nu}$ calculated around the critical point. The
         vertical line indicates the $\beta$ value where we got the smallest
         ${\chi}^2$ for the fit. At this point both ${\beta}/{\nu}$
         and ${\gamma}/{\nu}$ take the Ising values.}
       \label{exp}
     \end{figure}

    Figure \ref{Scomp} shows the average cluster size S for our four lattices at
    different
    $\beta$ values. The curves clearly peak close to the physical transition,
    and they shift slightly to the right the bigger the lattice size is.
    The second step was to perform other simulations
    in the very narrow range of $\beta$ values where our transition
    seems to occur. 
    We performed a finite size scaling analysis of our data 
    in this smaller range.     Figure \ref{exp} shows the results of 
    our analysis.
    The two curves represent the best fit values for the ratios ${\beta}/{\nu}$
    and ${\gamma}/{\nu}$ calculated for different beta values in the range
    between $\beta=3.410$ and $\beta=3.457$. We found the best ${\chi}^2$ at 
    $\beta=3.4407$, where ${\beta}/{\nu}$ and ${\gamma}/{\nu}$ take 
    the Ising values $1/8$ and $7/4$, respectively.

\section{CONCLUSIONS}

      We have shown that the deconfinement transition for SU(2) 
      gauge theory in 2+1 dimensions can be described in terms of
      Polyakov Loop percolation.


\begin{thebibliography}{9}

\bibitem{Co80} A. Coniglio, W. Klein, J.Phys. A: Math. Gen. {\bf 13}, 
  2775-2780 (1980). 

\bibitem{Stauf94} D. Stauffer and A. Aharony, 
  {\it Introduction to Percolation Theory}, 
      Taylor {\&} Francis, London 1994. 

\bibitem{Syk76} M. F. Sykes and D. S. Gaunt, J.Phys. A: Math. Gen. {\bf 9}, 
   2131-2137 (1976).

\bibitem{Krum74} H. M{\"u}ller-Krumbhaar, Phys. Lett. {\bf 48 A} , 459 (1974).

\bibitem{Svet82} B. Svetitsky and L. Yaffe, 
      Nucl. Phys. {\bf B210} [FS6], 423 (1982); L. Yaffe and
      B. Svetitsky, Phys. Rev. {\bf D26}, 963 (1982). B. Svetitsky Phys. Rep. 132,
      {\bf 1} (1986). 

\bibitem{Kar84} F. Green, F. Karsch, Nucl. Phys. {\bf B238}, 297-306 (1984).

\end{thebibliography}
\end{document}